\documentclass[final,1p,]{elsarticle}

\usepackage{lineno,hyperref}
\modulolinenumbers[5]

\journal{Astronomy and Computing}

\biboptions{authoryear}

\usepackage{jnl_abbrev}
\usepackage{amsmath}
\usepackage{tikz}
\newcommand{\checkmark}{\tikz\fill[scale=0.4](0,.35) -- (.25,0) -- (1,.7) -- (.25,.15) -- cycle;} 
\begin{document}

\begin{frontmatter}

\title{Estimation of Exoplanetary Planet-to-Star Radius Ratio with Homomorphic Processing}

\author[UCH_E]{Rodrigo Mahu}
\author[UCH_A]{Patricio Rojo}
\author[USACH]{Ali Dehghan Firoozabadi}
\author[USACH]{Ismael Soto}
\author[ESO]{Elyar Sedaghati}
\author[UCH_E]{Nestor Becerra Yoma\corref{mycorrespondingauthor}}
\cortext[mycorrespondingauthor]{Corresponding author}
\ead{nbecerra@ing.uchile.cl}

\address[UCH_E]{Department of Electrical Engineering, Universidad de Chile, Av. Tupper 2007, Santiago, Chile }
\address[UCH_A]{Departamento de Astronomia, Universidad de Chile, Camino el Observatorio 1515, Las Condes, Santiago, Chile}
\address[USACH]{Electrical Engineering Dept., Universidad de Santiago de Chile, Ave. Ecuador 3519, Zip Code: 9170124, Santiago, Chile}
\address[ESO]{European Southern Observatory, Alonso de Cordova 3107, Santiago, Chile}

\begin{abstract}

In this paper a homomorphic filtering scheme is proposed to improve the estimation of the planet/star radius ratio in astronomical transit signals. The idea is to reduce the effect of the short-term earth atmosphere variations. A two-step method is presented to compute the parameters of the transit curve from both the unfiltered and filtered data. A Monte Carlo analysis is performed
by using correlated and uncorrelated noise to determine the parameters of the 
proposed FFT filter. The method is tested with observations of WASP-19b and WASP-17b obtained with the FORS2 instrument at the Very Large Telescope (VLT).
The multi parametric fitting and the associated errors are obtained with the JKTEBOP
software. The results with the white light of the exo-planet data  mentioned above suggest that the homomorphic filtering can lead to substantial relative reductions in the error bars as high as 45.5\% and 76.9\%, respectively. The achieved reductions in the averaged error bars per channel were  48.4\% with WASP-19b and 63.6\% with WASP-17b. Open source MATLAB code to run the method proposed here can be downloaded from \url{http://www.cmrsp.cl}. This code was used to obtain the results presented in this paper.

\end{abstract}


\end{frontmatter}


\section{Introduction}
\label{intro}

Astronomical observations have greatly benefited from the introduction of digital detectors. The migration from analogue photographic-plate detectors to digital array of pixels such as the classic charge-coupled-devices (CCDs) happened towards the end of the previous century. The much enhanced sensitivity and robustness of the digital devices permitted giant leaps in observational astronomy. However, data analysis in astronomy may still greatly benefit from signal processing techniques.

The first exoplanet around a main sequence star was detected two decades ago \citep{mayor1995jupiter}. 
In the short period since that discovery, the field of exoplanetary research has evolved beyond detection, where atmospheric and interior characterization of these alien worlds have been made. A multi-wavelength time-series observation of an exoplanetary transit
with sufficient baseline allows for the measurement of the transmission signature
of stellar light's tangential path along the planetary terminator; likewise, an
equivalent observation of an exoplanetary occultation facilitates the measurement of the emission spectrum from planet's dayside. For both complementary
methods, the depth of the feature in the lightcurve is the parameter
that, after adequate modeling, relates to the exoplanetary atmospheric
properties like composition, temperature-pressure profiles
e.g. \citep{madsea09,forshaetal10,irodem10}; in the case of
transits, the depth is related to the planet-to-star
radius ratio.

Given the extra distortion by the Earth's atmosphere (telluric
effects), early exoatmosphere characterization was possible only
through space-based instrumentation, e.g. \citep{chabroetal02,viddesetal04,vidlecetal03} by detecting wavelength-dependent variations of transit depth in a time-series spectral observations as the planet crosses in front of the host star. More troublesome became securing the
first ground-based observations; after several null results
e.g. \citep{esp:ricdemetal03,esp:rojo06} the first bona-fide
detections of exoatmospheres were only possible in 2008
\citep{esp:redendetal08,esp:snealbetal08}.  Those first detections
observed the single target star correcting the telluric effects by
out-of-transit baselines, an approach that later kept yielding
detection of new atomic species as in \citep{esp:astroj13}.

However, simultaneously observing one or more reference stars together
with the exoplanetary system has been the favored method that,
complementing space-based measurements, have shown the large diversity
of exoplanetary atmospheres \citep[among others]{beadesetal11,mansouetal14,jorespetal13}.  Other ground-based methods with mixed
success involve searching for the planetary spectral lines as they
Doppler-wobble with respect to the telluric lines \citep{barbaretal10,rodloprib12,rodkurbar13,cubrojfor13}.

In this paper, we will study the effect on the measurement of
radius-ratio in an exoplanetary lightcurve after a filtered
homomorphic process. Section \ref{homo} presents the model, Section
\ref{valid} applies the model to a  real exoplanetary dataset for
validation, Section \ref{result} presents the results, and Section \ref{conc} presents the concluding remarks.

\section{Homomorphic Processing\label{homo}}

The main motivation of homomorphic processing is to separate signals
that have been combined not additively but instead through, for
instance, convolution or multiplication. This is achieved by mapping
this original system onto a different space. For instance, homomorphic
deconvolution and cepstral (frequency analysis in logarithmic
  space) analysis have been applied successfully in a variety of areas
including speech and audio processing, geophysics, radar, medical
imaging, and others \citep[among others]{oppenheim2004frequency,oppenheim1969speech,sreenivasan2015nonparametric}.

\begin{figure}
\includegraphics[width=\textwidth]{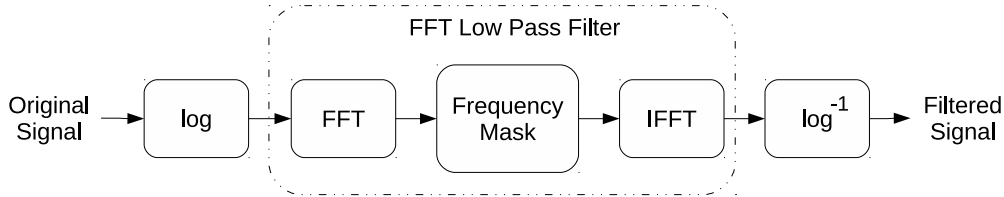}
\caption{Block diagram of the proposed homomorphic filtering technique
  to improve the estimation accuracy of the radius-ratio}
\label{fig:FFT_filter}
\end{figure}

In this paper we apply homomorphic processing to separate the
flux from the star of the target exoplanetary system and a simultaneously-observed
flux from a reference star in the same field of view, and later filter out some of the
telluric-induced high frequency components from the data. By doing so,
we seek to improve the accuracy of the exoplanet radius-ratio
estimation.  Figure \ref{fig:FFT_filter} shows the
proposed homomorphic filtering scheme.

\subsection{Model\label{homo:mod}}

The observed flux from the star with the transiting planet ($F_T$) for each $\lambda$ (wavelength) is modeled as follows:

\begin{equation} \label{eq:ft}
F_T(\lambda,t)=F_{TO}(\lambda,t)\times A_E(\lambda,t)+n_T(\lambda,t)
\end{equation}

\noindent and the corresponding observed flux for the reference ($F_R$) star can
expressed as:

\begin{equation} \label{eq:fr}
F_R(\lambda,t)=F_{RO}(\lambda)\times A_E(\lambda,t)+n_R(\lambda,t)
\end{equation}

\noindent where $F_{TO}(\lambda,t)$ and $F_{RO}(\lambda)$ denote the original
flux without the telluric  effects for the target and
reference star, respectively; $A_E(\lambda,t)$ corresponds to the
telluric atmosphere response; and, $n_T(\lambda,t)$ and
$n_R(\lambda,t)$ indicate the observation noise for the target and
reference stars.  Also, the original flux from the reference star,
$F_{RO}(\lambda)$, is independent of time.

The transit light curve for a given $\lambda$,
$F_{TO}(\lambda,t)$, is described by the Mandel \& Agol 
model \citep{mandel2002analytic}.  The flux received from the target star depends on the radiated
energy, which is a function of $\lambda$, and on the exoatmosphere
absorption during the planetary transit that also depends on
$\lambda$.  In the case of the reference star we can assume that there
is no time dependency.

The telluric effect on the  flux received from the target and
reference stars, $A_E(\lambda,t)$, is the standard attenuation term
($\propto e^{-\tau}$). This component is wavelength and time dependent
as the atmosphere changes.  Atmosphere perturbations can be classified
as long- or short-term. The air mass changes slowly in a few-hour
observation window as the star moves in the sky. In contrast,
parameters such as density, pressure and amount of water may present
rapid fluctuations on timescales of 10 or 15 minutes.

The additive observation noise is composed of a white and a red noise
component. Both types of noise present high frequency energies,
although they are more significant in the white noise.

\subsection{Mapping\label{homo:map}}

We study homomorphic mapping into logarithmic space by first applying
the logarithm operator to the observed light fluxes in
(\ref{eq:ft}) and (\ref{eq:fr}) to transform the multiplication to
addition:
             
\begin{equation} \label{eq:lft}
\begin{split}
\log \left[F_T(\lambda,t) \right]= &\log\left[F_{TO}(\lambda,t)\right] + \log\left[A_E(\lambda,t)\right]\\ 
& +\log\left[1+ \frac{n_T(\lambda,t)}{F_{TO}(\lambda,t)A_E(\lambda,t)}\right]
\end{split}
\end{equation}                     

\begin{equation} \label{eq:lfr}
\begin{split}
\log \left[F_R(\lambda,t) \right]= &
\log\left[F_{RO}(\lambda)\right] + \log\left[A_E(\lambda,t)\right]\\ 
& +\log\left[1+ \frac{n_R(\lambda,t)}{F_{RO}(\lambda)A_E(\lambda,t)}\right]
\end{split}
\end{equation}        

By considering that $F_{TO}(\lambda,t)A_E(\lambda,t)\gg
n_T(\lambda,t)$ and $F_{RO}(\lambda)A_E(\lambda,t)\gg n_R(\lambda,t)$,
we can make use of the approximation $\log(1+x)\approx x$, if $x\ll
1$.  The normalization of $F_T(\lambda,t)$ with respect to
$F_R(\lambda,t)$, i.e. the subtraction of (\ref{eq:lfr}) from
(\ref{eq:lft}), can then be expressed as:

\begin{equation} \label{eq:lratio}
\begin{split}
\log \left[\frac{F_T(\lambda,t)}{F_R(\lambda,t)} \right]= & \log\left[F_{TO}(\lambda,t)\right] - \log\left[F_{RO}(\lambda)\right] \\ &+ \frac{n_T(\lambda,t)}{F_{TO}(\lambda,t)A_E(\lambda,t)} - \frac{n_R(\lambda,t)}{F_{RO}(\lambda)A_E(\lambda,t)}
\end{split}
\end{equation} 

\noindent where the component $\log\left[A_E(\lambda,t)\right]$ in
(\ref{eq:lft}) and (\ref{eq:lfr}) are canceled out since the target and the reference star's angular separation on the sky is negligible.

\subsection{Filtering\label{homo:filt}}

The flux ratio (\ref{eq:lratio}) is thus processed with a
low-pass Fast Fourier Transform (FFT) filter to suppress or reduce the high frequency
components of the noise terms. Basically, the main idea is to reduce
the distortion caused by the short term variations of the telluric
response and of the observation noise. If the attenuated components
are faster than the transit time, then a more accurate estimation of
the radius ratio could be achieved.

\begin{figure}
\includegraphics[width=\textwidth]{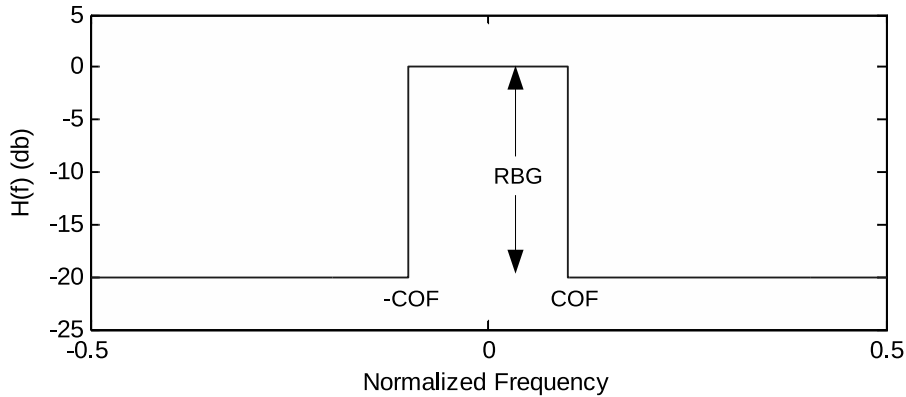}
\caption{FFT Low-pass filter mask defined by the cut-off frequency (COF) and the rejection band gain (RBG).}
\label{fig:FFT_mask}
\end{figure}

The low-pass filter is applied in the frequency domain to prevent the
addition of delay and phase distortion of the frequency components.
The resulting filter is non-causal and requires processing the whole signal
at once. This is achieved by multiplying the FFT spectrum with a
frequency mask that attenuates all the components above a cut off
frequency. Figure \ref{fig:FFT_mask} shows the low-pass filter mask
defined by the cut-off frequency (COF) and the rejection band gain
(RBG) in dB. Then, the inverse FFT is employed to come back to the
time domain. Finally, after low-pass filtering the log of the
normalized transit flux curve, the inverse logarithm operator is
applied.

\subsection{Parameter estimation\label{parameters}}

We follow the standard prescription of  \citet{mandel2002analytic} to
find the best-fit parameters to the data.  Since for characterizing
extrasolar atmospheres, it is typically required to have a precision
better than $10^{-3}$ of the incoming flux.  It is important to verify
that new methods do not add unwanted systematic effects that could
offset the measurement. In fact, our early tests did show that a
low-pass filtering distorts the shape of the lightcurve for the
estimation of some of the parameters (most greatly for the sum of radii
and the  inclination).

\begin{figure}
\centering
\includegraphics[width=\textwidth]{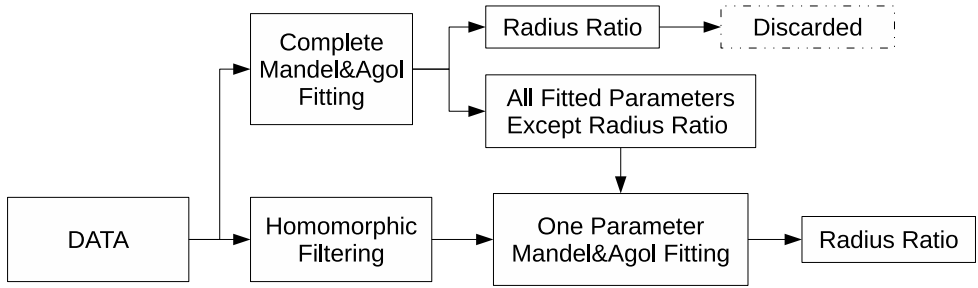}
\caption{Diagram for the two-step method of estimation.}
\label{fig:MC2Step}
\end{figure}

Therefore, in order to counteract the above-mentioned distortion
induced by the homomorphic low-pass filter, and since the most
relevant parameter for exoatmospheric measurements is the
radius-ratio, we develop a two-step method (Fig. \ref{fig:MC2Step})
that only tries to fit this parameter, while minimizing errors associated to
possible degeneracies with other parameters. First, the observed transit curve is fitted with the Mandel \& Agol
model. In this paper this is achieved by using JKTEBOP \citep{southworth2013solar}. As a result the fitted parameters are estimated and the radius ratio coefficient is discarded. Second, the homomorphic low-filtering is applied and the resulting curve is also fitted with the Mandel \& Agol model. However, in contrast to the previous step, the Mandel \& Agol fitting makes use of all the parameters previously estimated except radius ratio, which is now determined
(Table
\ref{fittable}).

\begin{table}
  \begin{center}
  \caption{\label{fittable}Parameters from the 
    \citet{mandel2002analytic} prescription that are fit during the
    two-step method}
  \begin{tabular}{l|c|c}
    \hline\hline
    Parameter & \multicolumn{2}{c}{Fit to curve} \\
    Name & unfiltered & filtered \\
    \hline
    radius ratio & \checkmark & \checkmark\\
    radii sum  & \checkmark & \\
    inclination & \checkmark & \\
    linear limb darkening & \checkmark & \\
    quadratic limb darkening & \checkmark & \\
    period  &  \checkmark & \\
    mid transit & \checkmark & \\
    \hline\hline
  \end{tabular}
  \end{center}
\end{table}

\section{Validation\label{valid}}

\subsection{The Data\label{valid:data}}

\begin{figure}
\centering
\includegraphics[width=\textwidth]{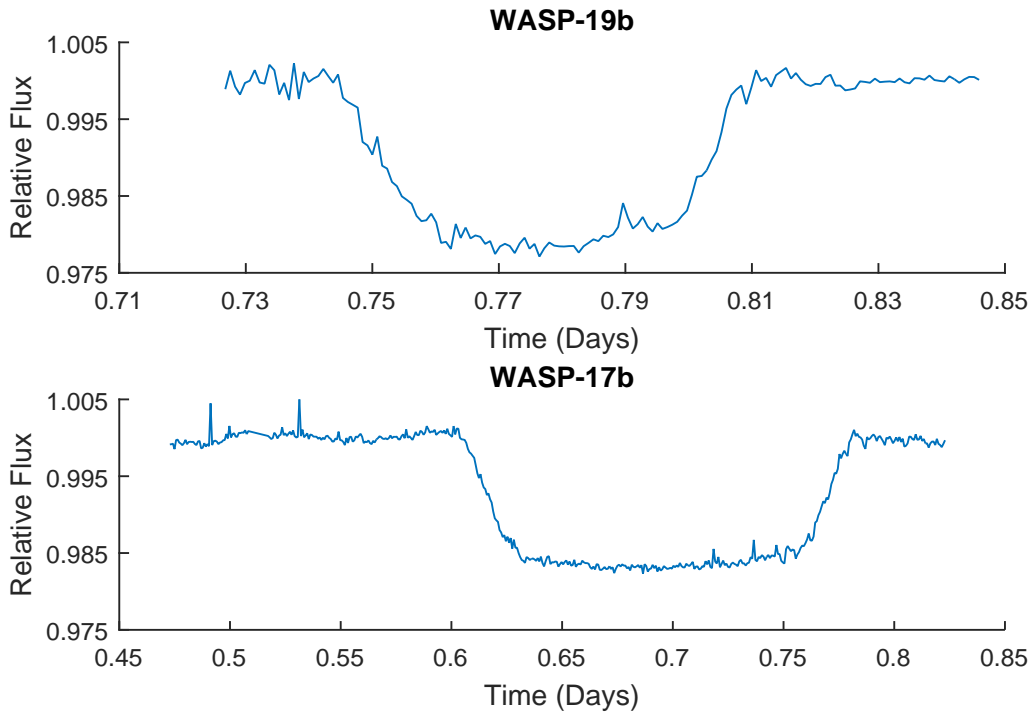}
\caption{White light data from WASP-19 (top) and WASP-17 (bottom).  The curves were obtained by summing all the spectral channels from 535nm to 837nm and from 570nm  to 790nm for WASP-19b and WASP-17b, respectively. Finally, the curves were normalized by the flux average of the reference star.}
\label{fig:RealSignal}
\end{figure}

We evaluated the proposed method with the transit time series corresponding to planets WASP-19b \citep{sedaghati2015regaining} and WASP-17b \citep{sedaghati2016potassium}. The WASP-19b  and WASP-17b data were obtained on the nights of Nov 16, 2014, and Jun 18, 2015, respectively, using FORS2 at the VLT with Grism 600RI.
In the case of WASP-19b, 30''$\times$10'' slits were used on six reference stars, in addition to WASP-19, and the exposure time was 30s. For WASP-17b, 30''$\times$15'' slits were employed  on five reference stars, in addition to WASP-17, and the exposure time was 35s
(Fig.\ \ref{fig:RealSignal}).

\subsection{Monte Carlo-based estimation method\label{valid:mc}}

\begin{figure}
\includegraphics[width=\textwidth]{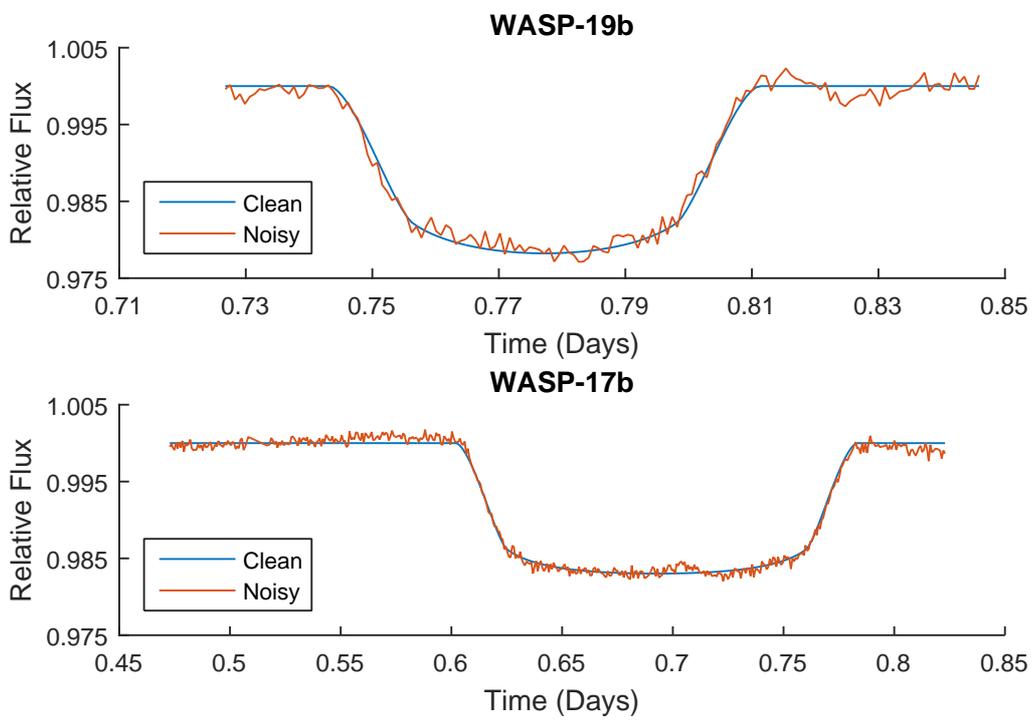}
\caption{Examples of artificially generated noisy transit curves employed in the Monte Carlo simulations: WASP-19b (top) and WASP-17b (bottom).}
\label{fig:ArtificialSignal}
\end{figure}

To estimate the parameters of the proposed homomorphic FFT filter, we generated a set of artificial 1000 transit noisy signals for each planet considered here, i.e. WASP-19b and WASP-17b. This dataset was produced by adopting the following procedure: first, a synthetic transit curve was obtained for each planet with the Mandel \& Agol model by making use of the  estimated transit parameters of WASP-19b and WASP-17b  according to \citep{sedaghati2015regaining} and \citep{sedaghati2016potassium}, respectively; then, white noise and red noise  were added to the synthetic transit curves.   The red noise used here presents a spectral density distribution that decreases at a rate of $f^{-2}$ \citep{carter2009parameter}. The red and white noise power were  the same. The resulting SNR after the addition of both types of noise was equal to the SNR of the white light curves shown in Figure \ref{fig:RealSignal}.  Figure \ref{fig:ArtificialSignal} depicts example of the noisy transit curves generated for WASP-19b and WASP-17b. The grid search described in Subsection \ref{valid:fo} was employed to estimate the homomorphic FFT filter parameters, i.e. COF and RBG.

\begin{figure}
\includegraphics[width=\textwidth]{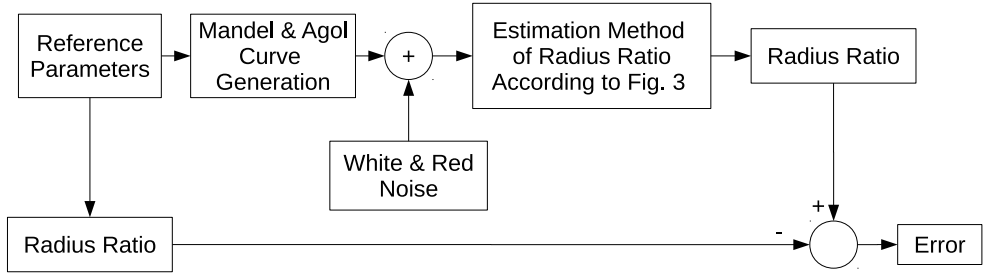}
\caption{Monte Carlo simulation to estimate the parameters of the homomorphic FFT filter.}
\label{fig:Montecarlo}
\end{figure}

Figure \ref{fig:Montecarlo} shows the Monte Carlo-based estimation method. The 1000 noisy signals
were filtered with the homomorphic low-pass filter. Then JKTEBOP was
used to estimate the transit parameters for the unfiltered and
filtered signals.  After estimating the parameters with the original
1000 unfiltered and filtered signals, we calculated the average
estimation of radius ratio and corresponding error bar considering an interval of
68\% of confidence that corresponds to one gaussian standard deviation.

\begin{figure}
\includegraphics[width=\textwidth]{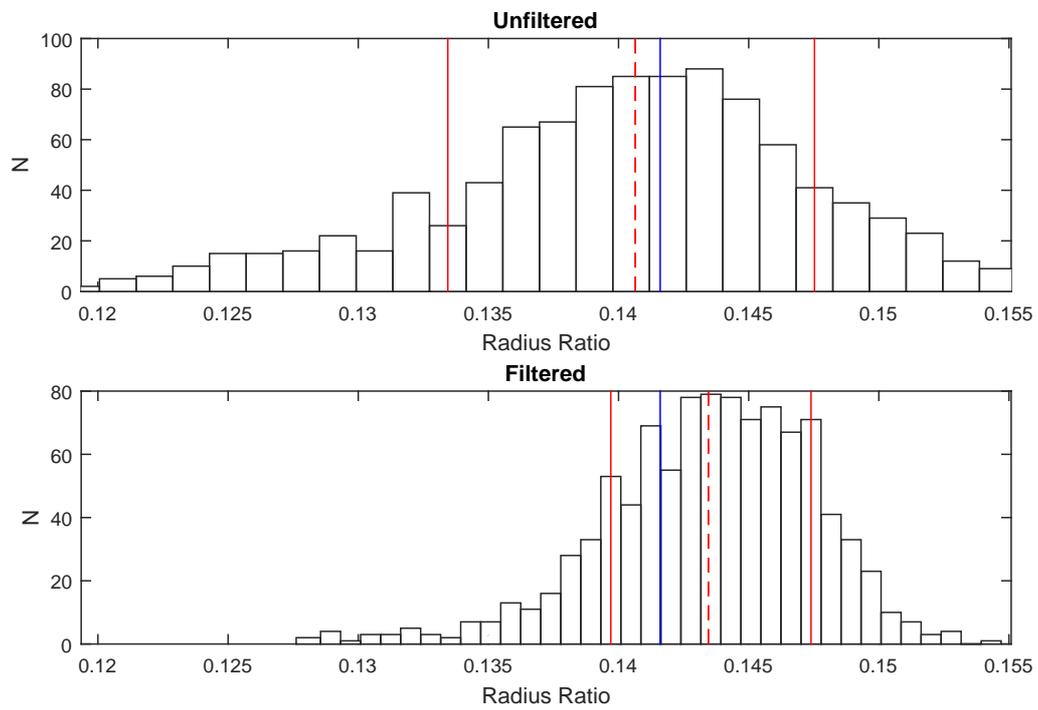}
\caption{
Histograms of the estimated radius-ratio for WASP-19b with the early test: unfiltered (top) and filtered
(bottom) data. The red continuous lines, at the outer sides of the histogram, bound the 68\% confidence interval. The blue and
dashed red lines, in the middle of the histogram, indicate the reference radius-ratio and the average estimation, respectively. (For interpretation of the references to colour in this figure legend, the reader is referred to the web version of this article.)
}
\label{fig:Hist1Step}
\end{figure}

As can be seen in Figure \ref{fig:Hist1Step}, in the early test (i.e. fitting all parameters after the filter), the homomorphic low-pass filter led to a reduction of 45.4\% in the estimation of error bar for the radius ratio with WASP-19b.
However, the method is unable to find the correct estimation for inclination and radii sum, 
and the error bar of the mid transit increased, because of the distortion incorporated by the homomorphic low-pass
filter in the transit curve transition.

\begin{figure}
\includegraphics[width=\textwidth]{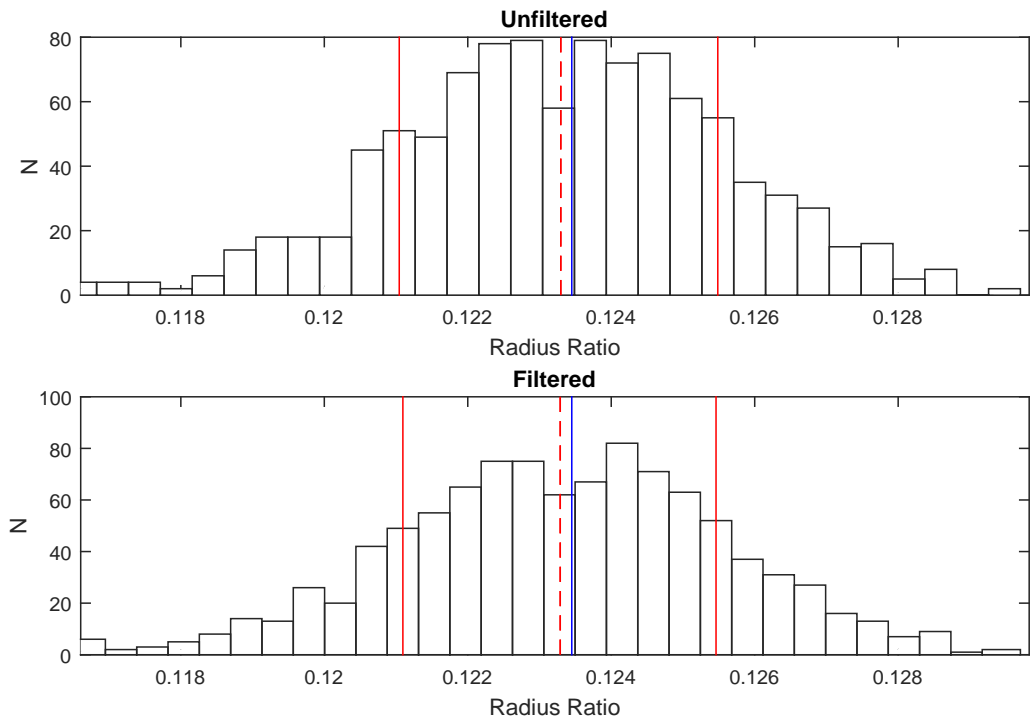}
\caption{
Histograms of the estimated radius-ratio for WASP-17b with the early test: unfiltered (top) and filtered
(bottom) data. The red continuous lines, at the outer sides of the histogram, bound the 68\% confidence interval. The blue and
dashed red lines, in the middle of the histogram, indicate the reference radius-ratio and the average estimation, respectively.  (For interpretation of the references to colour in this figure legend, the reader is referred to the web version of this article.)}
\label{fig:Hist1StepWasp17}
\end{figure}

As can be seen in Fig. \ref{fig:Hist1StepWasp17}, the early test with WASP-17b leads to a low error reduction of 1.69 \%, which is much lower than the error reduction achieved with WASP-19b. This result must be due to the fact that, according to Table \ref{SNRtotal},  the white light curve of WASP-17b provides an SNR that is 77\% higher than the one observed in the WASP-19b light curve when the noise signal is obtained   as the  residual of the JKTEBOP estimation. In contrast, Table \ref{SNRtotal} also shows that the SNR averaged across all the channels is just 20\% higher in WASP-17b than in WASP-19b. This apparent contradiction is explained as follows. According to Table  \ref{Noise_comp}, the noise in WASP-17b is much less correlated between channels than in WASP-19b. Consequently, when the residual noise signals are summed over all the channels, the resulting noise energy is much lower with WASP-17b than with WASP-19b. Coefficient Q in Table \ref{Noise_comp} is defined as the quotient of the energy of the signal resulting from the summation of all the channel noises by the summation of the individual channel noise energies as in Eq. \ref{eq:Qratio}  (see Table \ref{SNRchanels}):

\begin{equation} \label{eq:Qratio}
Q=\frac{\sum_t \left[ Noise(t)\right] ^2}{\sum_\lambda \sum_t  \left[Noise(t,\lambda)\right] ^2}
\end{equation}

\noindent where $Noise(t,\lambda)$ denote sample $t$ of the noise signal at channel $\lambda$ and,

\begin{equation} \label{eq:Noise}
Noise(t)= \sum_\lambda Noise(t,\lambda)
\end{equation}

\noindent  Accordingly, WASP-17b delivers a Q value that is 3.5 times lower than the one obtained with WASP-19b.

\begin{figure}
\centering
\includegraphics[width=\textwidth]{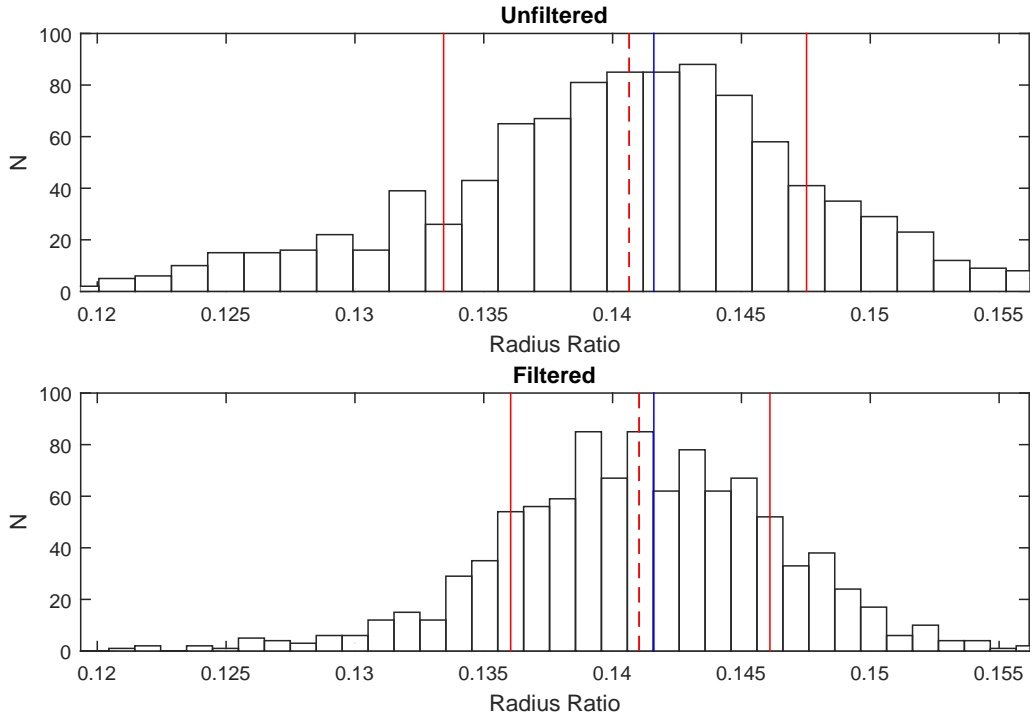}
\caption{Histograms of the estimated radius ratio for WASP-19 with the proposed estimation method: unfiltered
(top) and filtered (bottom) data. The red continuous lines, at the outer sides of the histogram, bound the 68\% confidence
interval. The blue and dashed red lines, in the middle of the histogram, indicate the reference radius-ratio and the average
estimation, respectively. (For interpretation of the references to colour in this figure legend, the reader is referred to the web version of this article.)
}
\label{fig:Hist2Step}
\end{figure}

\begin{figure}
\centering
\includegraphics[width=\textwidth]{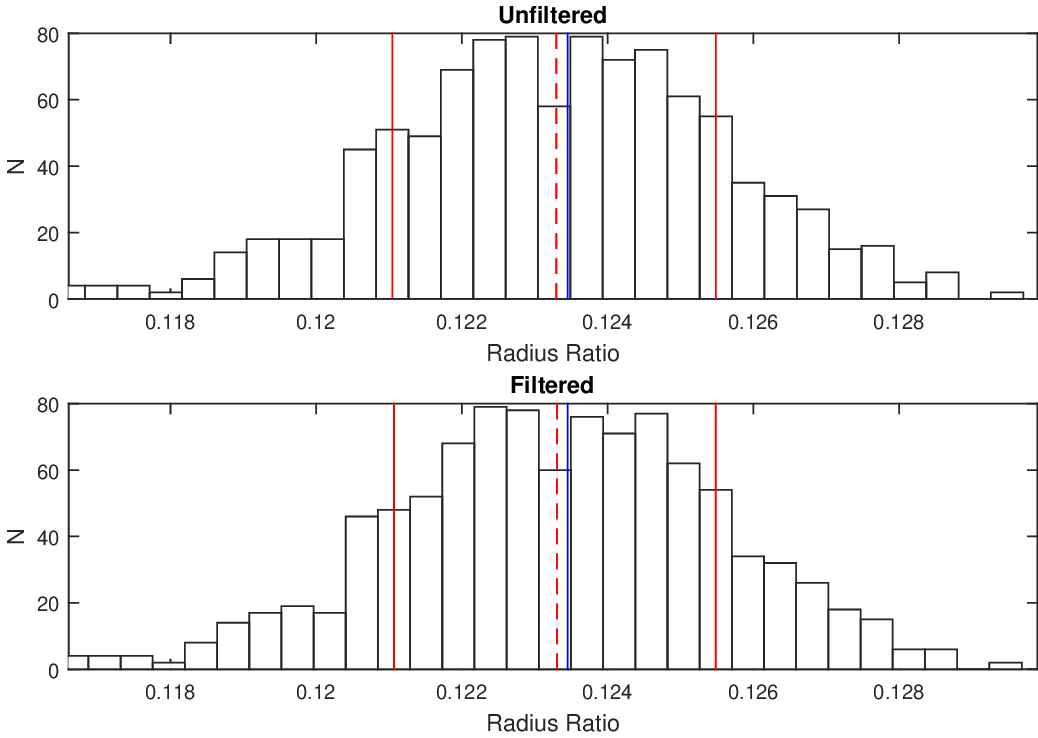}
\caption{Histograms of the estimated radius ratio for WASP-17 with the proposed estimation method: unfiltered
(top) and filtered (bottom) data. The red continuous lines, at the outer sides of the histogram, bound the 68\% confidence
interval. The blue and dashed red lines, in the middle of the histogram, indicate the reference radius-ratio and the average
estimation, respectively. (For interpretation of the references to colour in this figure legend, the reader is referred to the web version of this article.)
}
\label{fig:Hist2StepWasp17}
\end{figure}

Results are shown in Fig. \ref{fig:Hist2Step} for the proposed two-step method. The reduction in the estimation error bar for WASP-19 was as high as 28.4\% without additional distortion. Fig. \ref{fig:Hist2StepWasp17}  shows an improvement of 0.66\% for the case of WASP-17.

\begin{table}
  \begin{center}
  \caption{\label{SNRtotal} SNR in the white light curves and the SNR averaged over all the channels with WASP-17b and WASP-19b. The noise is estimated as the residual between the real signal and the JKTEBOP estimation of the transit curve.}
  \begin{tabular}{c|c|c}
    \hline\hline
    Planet & White light curve &	SNR averaged over  \\
    Name & SNR &  all the channels	 \\
    \hline
    WASP-17b &	147.99 &	74.75 \\
    WASP-19b &	83.45 &	62.30\\
    
    \hline\hline
  \end{tabular}
  \end{center}
\end{table}

\begin{table}
  \begin{center}
  \caption{\label{SNRchanels} SNR at each channel with WASP-17b and WASP-19b. The noise is estimated as the residual between the real signal and the JKTEBOP estimation of the transit curve. } 
  \begin{tabular}{c|c||c|c}
    \hline\hline
    \multicolumn{2}{c}{ WASP 17b} & \multicolumn{2}{c}{ WASP 19b} \\
    Channel & SNR &	Channel & SNR  \\
    \hline
		&			&	560.00	nm	&	57.02		\\
570.00	nm	&	12.25		&	570.00	nm	&	52.57		\\
580.00	nm	&	27.07		&	581.00	nm	&	53.21		\\
590.00	nm	&	63.05		&	590.00	nm	&	58.70		\\
600.00	nm	&	83.91		&	601.00	nm	&	61.17		\\
610.00	nm	&	94.11		&	610.00	nm	&	65.24		\\
620.00	nm	&	82.24		&	620.00	nm	&	67.52		\\
630.00	nm	&	79.32		&	630.00	nm	&	64.09		\\
640.00	nm	&	85.83		&	640.50	nm	&	60.31		\\
650.00	nm	&	95.70		&	650.00	nm	&	61.87		\\
660.00	nm	&	99.10		&	660.50	nm	&	62.94		\\
670.00	nm	&	111.27		&	670.00	nm	&	63.95		\\
680.00	nm	&	118.00		&	681.25	nm	&	67.56		\\
690.00	nm	&	99.97		&	690.00	nm	&	69.80		\\
700.00	nm	&	93.96		&	701.25	nm	&	64.27		\\
710.00	nm	&	98.89		&	711.25	nm	&	57.97		\\
720.00	nm	&	83.70		&	720.00	nm	&	57.44		\\
730.00	nm	&	81.65		&	731.25	nm	&	49.19		\\
740.00	nm	&	84.22		&	740.00	nm	&	59.50		\\
750.00	nm	&	60.62		&	748.00	nm	&	73.07		\\
760.00	nm	&	34.58		&	760.00	nm	&	45.15		\\
770.00	nm	&	34.16		&	768.00	nm	&	51.08		\\
780.00	nm	&	47.20		&	780.00	nm	&	74.46		\\
790.00	nm	&	48.41		&	790.00	nm	&	72.17		\\
		&			&	800.00	nm	&	73.82		\\
		&			&	810.00	nm	&	71.10		\\
		&			&	820.00	nm	&	66.99		\\
    
    \hline\hline
  \end{tabular}
  \end{center}
\end{table}

\begin{table}
  \begin{center}
  \caption{\label{Noise_comp} Correlation coefficient between channel noises averaged across all the channels and metric Q according to (\ref{eq:Qratio}) with WASP-17b and WASP-19b.}
  \begin{tabular}{c|c|c}
    \hline\hline
    Comparison & \multicolumn{2}{c}{Planet Name} \\
    Criteria & WASP-17b &	WASP-19b \\
    \hline
    Average correlation coefficient & & \\ 
    of the noise between channels. &	0.4521 & 0.7309	 \\
    Q, as defined in (\ref{eq:Qratio}). & 0.0431	 &	0.1513 \\
    
    \hline\hline
  \end{tabular}
  \end{center}
\end{table}

\subsection{Filter optimization \label{valid:fo}} 	

The homomorphic FFT low-pass filter defined in Figs. \ref{fig:FFT_filter} and \ref{fig:FFT_mask} has two parameters that need to be tuned, COF and RBG. These parameters were optimized by making use of a two-dimensional grid search methodology. In the case of WASP-19b, 153 samples, the 256 FFT was obtained and the optimal COF was searched in the integer interval [1, 127]. The optimal RBG is found within the set [10dB, 20dB, 30dB, 40dB, 50dB, 60dB, 70dB, 80dB]. For each pair (COF, RBG) the Monte Carlo simulation, described in Subsection \ref{valid:mc} , was performed and the optimal pair (COF, RBG) was determined with respect to the lowest interval of 68\% confidence that includes the correct radius ratio. Accordingly, the optimal filter parameters for WASP-19b are: for the early test, COF equal to five  FFT samples and RBG equal to 10dB; and, for the proposed method, COF corresponds to 14 FFT samples and RBG is equal to 30dB. The WASP-17b data contains 477 samples and the 512 FFT was computed. To adopt a similar resolution to WASP-19b, the homomorphic FFT low-pass filter for WASP-17b was optimized by searching COF in the interval [1, 255] and discarding one every two FFT samples. Accordingly, the optimal filter parameters for WASP-17b are: for the early test, COF equal to 159  FFT samples and RBG equal to 40dB; and, for the proposed method, COF corresponds to 15 FFT samples and RBG is equal to 10dB.

\subsection*{Distortion analysis}

\begin{figure}
\centering
\includegraphics[width=\textwidth]{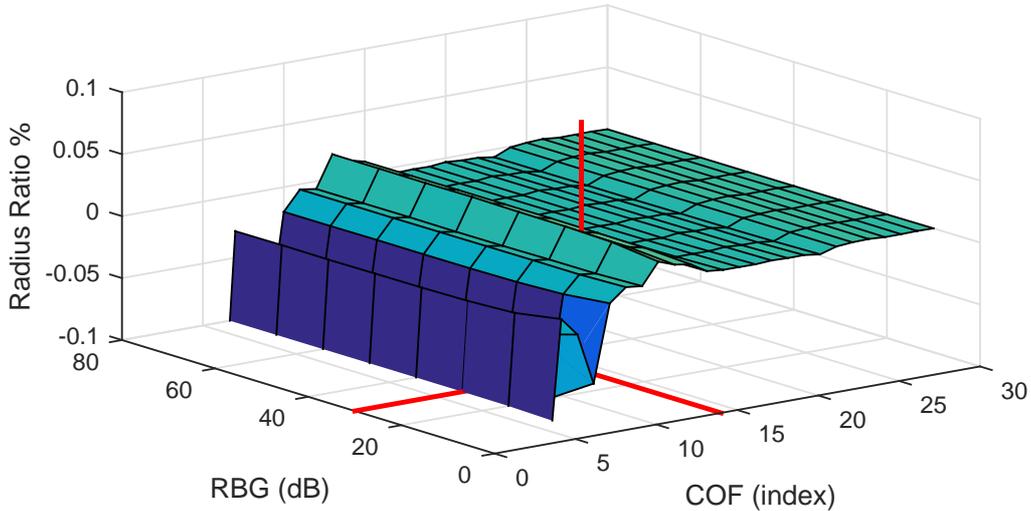}
\caption{Evaluation of the distortion introduced in the original clean transit curve by the
proposed method. The error curves were obtained when radius-ratio was made equal to
0.1416. The filtered radius-ratios were estimated with JKTBOP. The optimal filter (COF=14 FFT samples and RBG=30 dB) is indicated by the red lines.(For interpretation of the references to colour in this figure legend, the reader is referred to the web version of this article.) }
\label{fig:Sensibility}
\end{figure}

The distortion introduced by the proposed method and its dependence on filter parameters was evaluated by processing the clean
transit curves with the scheme shown in Fig.\ref{fig:FFT_filter} by applying the
optimal FFT low-pass filter with COF=14 FFT samples and RBG=30dB. The clean
transit curves were generated by using the parameters from WASP-19b and varying
radius-ratio. The procedure was repeated with six radius-ratios  representative of real exoatmospheric modulation along a typical spectrum: 0.136, 0.138, 0.14, 0.1416,
0.142 \& 0.144. Then, the error between the radius-ratios in the original clean and filtered
transit curves is computed. Figure \ref{fig:Sensibility} shows the error curve when radius-ratio was
made equal to 0.1416. As can be seen in Fig. \ref{fig:Sensibility}, there is a wide range of values for COF and
RBG, including the optimal ones, where the error is lower than $10^{ -4}$. The same behavior is
obtained with the other five reference radius-ratios. This result strongly suggests that the
proposed homomorphic filtering does not introduce any significant distortion into the
observed original transit curves

\section{Results \label{result}}

\begin{figure}
\centering
\includegraphics[width=\textwidth]{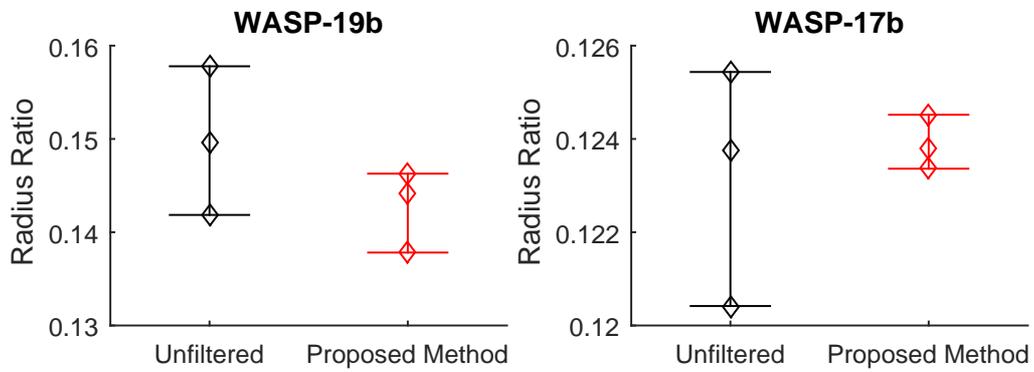}
\caption{Estimation of radius ratio on white light data: black, unfiltered;  and, red, with the proposed
method. (For interpretation of the references to colour in this figure legend, the reader is referred to the web version of this article.)}
\label{fig:ErrorBar}
\end{figure}

\begin{figure}
\centering
\includegraphics[width=\textwidth]{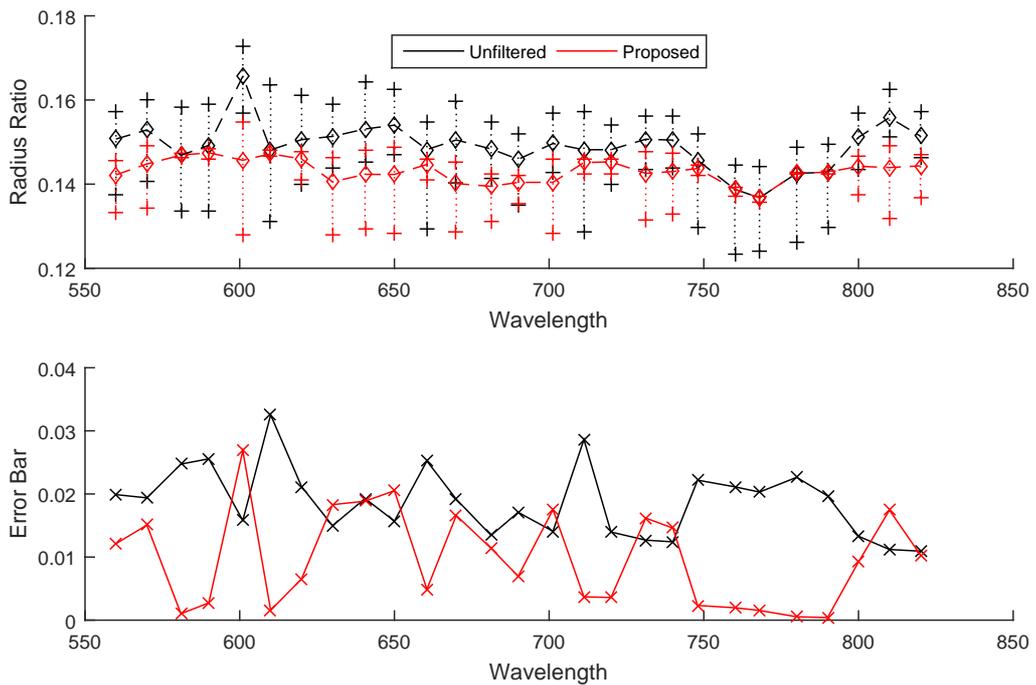}
\caption{Estimations of radius-ratio with the corresponding confidence intervals by employing
JKTEBOP: black, unfiltered data; red, proposed homomorphic filtering
method. The wavelength channels are defined in \citep{sedaghati2015regaining} for
WASP-19b. (For interpretation of the references to colour in this figure legend, the reader is referred to the web version of this article.)}
\label{fig:WASP19REAL}
\end{figure}

\begin{figure}
\centering
\includegraphics[width=\textwidth]{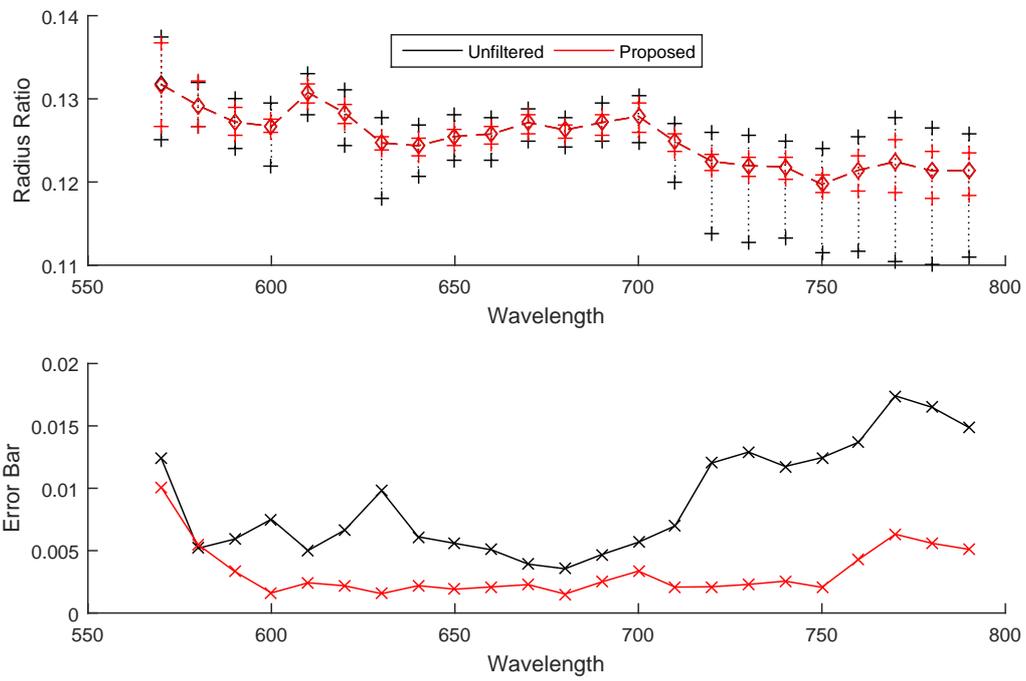}
\caption{Estimations of radius-ratio with the corresponding confidence intervals by employing
JKTEBOP: black, unfiltered data; red, proposed homomorphic filtering
method. The wavelength channels are defined in \citep{sedaghati2016potassium} for
WASP-17b. (For interpretation of the references to colour in this figure legend, the reader is referred to the web version of this article.)}
\label{fig:WASP17REAL}
\end{figure}

We used JKTEBOP with the white light from WASP-19b and WASP-17b to compare the ordinary estimation of the transit curve parameters (baseline) with the one obtained by making use of  the proposed method (see Fig. \ref{fig:ErrorBar}). According to Fig. \ref{fig:ErrorBar}, the radius ratios for WASP-19b and WASP-17b are consistent because there is a clear intersection between the error intervals obtained with the ordinary baseline processing and the one achieved with the proposed method. Nevertheless, the homomorphic filtering delivers  error intervals that are   45.5\% and 76.9\% smaller for WASP-19b and WASP-17b, respectively. 

Figure \ref{fig:WASP19REAL} shows the radius ratios computed by employing the proposed restricted estimation method with the 27 channels of WASP-19 reported in \citep{sedaghati2015regaining}. As can be seen in Fig. \ref{fig:WASP19REAL}, both sets of estimated radius ratios are also consistent but the proposed method provided error bars that are  48.4\% lower in average than without filtering.  A similar analysis results from  Figure \ref{fig:WASP17REAL} where the radius ratios obtained with the 23 channels of WASP-17, according to  \citep{sedaghati2016potassium},  were estimated with and without the proposed homomorphic filtering. As can be seen in Figure \ref{fig:WASP17REAL}, a dramatic  average reduction of 63.6\%  in the error bars was achieved with WASP-17b.

\section{Conclusions \label{conc}}

The homomorphic filtering technique proposed here is applied to improve the estimation of the radius ratio in astronomical transit signals. The main motivation is to reduce the interference from the short-term earth atmosphere variations. The parameters of the homomorphic FFT filter are estimated with a Monte Carlo-based scheme. The proposed method was tested with the white light of WASP-19b and WASP-17b real data using JKTEBOP, and was able to lead to dramatic reductions of 45.5\% and 76.9\% in the error bars, respectively.  Similar reductions in the averaged error bars per channel were achieved with WASP-19b and WASP-17b: 48.4\% and 63.6\%, respectively. Moreover, the sensibility analysis shows that the distortion caused by the technique is lower than the target precision, i.e. 0.1\%. Finally, the application of the homomorphic filtering scheme to additional exo-planet real data is proposed as a future research.

\section*{Acknowledgements}

The research leading to these results was funded by CONICYT-ANILLO 
project ACT 1120,
project POSTDOC\_DICYT (No.041613DA\_POSTDOC), Universidad de Santiago de Chile (USACH).

\section*{References}

\bibliography{References}

\end{document}